\documentclass[runningheads]{llncs}

\usepackage[T1]{fontenc}
\usepackage{graphicx}
\usepackage{amsmath,amssymb}
\usepackage{booktabs}
\usepackage{subcaption}
\usepackage{numprint}
\usepackage{xcolor}
\usepackage{hyperref}

\npthousandsep{\,}
\npdecimalsign{.}
\newcommand{\num}[1]{\numprint{#1}}
\newcommand{\pct}{\%}

\graphicspath{{figs/}}

\begin{document}

\title{Tuning Dispatch Thresholds for Fixed\\ Last-Mile Routes: A Simulation-Based\\ Pareto Analysis of a Production Policy}
\titlerunning{Tuning Dispatch Thresholds for Fixed Last-Mile Routes}

\author{Alexander Ponomarenko\inst{1} \and Ilya Antonov\inst{2}}
\authorrunning{A. Ponomarenko and I. Antonov}
\institute{Laboratory of Algorithms and Technologies for Network Analysis, HSE University,
Nizhny Novgorod, Russia\\
\email{aponomarenko@hse.ru}\\
\and
HSE University, Nizhny Novgorod, Russia\\
\email{a222i@ya.ru}
}

\maketitle

\begin{abstract}
Many parcel networks dispatch vehicles on \emph{fixed routes} using a simple
load-accumulation rule: a truck leaves the depot for a fixed route as soon as
the volume (or item count) waiting for that route crosses a threshold. The
threshold is usually parameterised as an affine function of route length,
$\tau_r=\beta+\gamma\,d_r$, and the pair $(\beta,\gamma)$ is chosen once and
frozen into production. This paper studies how good that frozen choice actually
is, treating the question as a data-intensive, data-driven decision-making
problem over a full month of real operational flow. Using a discrete-event
simulator that replays the recorded arrival stream and reconstructs every trip,
we sweep the $(\beta,\gamma)$ design space, evaluate the two competing
objectives---company operating cost and average parcel lead time---and recover
the Pareto frontier of efficient policies for two deployed variants
(volume-triggered and item-count-triggered). The two policies turn out to be in
strikingly different states of tune. The volume-threshold configuration lies on
its own Pareto frontier: the simulator finds no $(\beta,\gamma)$ pair that
strictly dominates it, so the deployed policy is \emph{already
Pareto-efficient}---an unusual positive audit result. The item-count
configuration is the opposite: it is dominated by a concrete simulated
configuration that is both faster and cheaper, and the available cost saving at
equal lead time is about \num{5.0}\,\pct{}. We trace the item-count policy's
inefficiency to a base that is too large and a length coefficient that is too
small for the deployed truck capacity, and show that a \emph{steeper}
threshold---lower base, higher slope---is preferable. Because the remedy is a
two-scalar reconfiguration, the analysis converts directly into an actionable,
zero-capital recurring saving.

\end{abstract}

\section{Introduction}
\label{sec:intro}

A large fraction of last-mile parcel volume in regional networks moves on
\emph{fixed routes}: ordered sequences of pickup-and-delivery offices (PUDO
points) that are fixed for a planning period, each served from a single depot.
Unlike on-demand routing, the open operational decision is not \emph{where} a
vehicle goes---the geometry is fixed---but \emph{when} it leaves. Dispatching too
early wastes capacity and inflates the number of trips (and therefore distance,
driver hours and cost); dispatching too late lets parcels age in the depot
buffer and inflates customer lead time. The operator therefore faces a genuine
two-objective trade-off between cost and speed, and the policy that resolves it
is encoded in a small number of configuration parameters that are typically set
once and rarely revisited.

In the network studied here the dispatch decision is governed by a threshold
rule. For each route $r$ the system accumulates the waiting load and releases a
truck when that load reaches a threshold
\begin{equation}
\tau_r=\beta+\gamma\,d_r ,
\label{eq:threshold}
\end{equation}
where $d_r$ is the round-trip length of route $r$ and $(\beta,\gamma)$ are two
scalars shared by all routes. Two variants are in operational use:
\texttt{flexible\_vol}, where the accumulated \emph{volume} triggers dispatch,
and \texttt{flexible\_shk}, where the accumulated \emph{item count}
(\emph{ShK}, scanned barcodes) triggers it. The pair $(\beta,\gamma)$ is
calibrated once and embedded in the production configuration. Because the rule
is so compact, the natural question is whether the deployed values are actually
efficient---or whether the same level of service could be bought more cheaply.

The two variants are not interchangeable in practice, and the operator's
historical choice between them was driven by data quality rather than by
optimality. In the operator's warehouse management system the per-parcel
volume tracks the physically loaded volume only approximately---stored volumes
deviate from the true value for a non-trivial fraction of parcels---whereas the
scanned-barcode count is recorded exactly at intake. A policy that meters
dispatch by item count is therefore both simpler to operate and less sensitive
to measurement noise, and \texttt{flexible\_shk} is the policy that was
originally deployed in production with $(\beta,\gamma)$ values set by hand. The
volume-triggered variant \texttt{flexible\_vol} is the natural alternative once
a reliable volume signal becomes available, and its production
$(\beta,\gamma)$ was set later. The audit reported here therefore serves three
complementary purposes: (i) verify whether the hand-tuned
\texttt{flexible\_shk} parameters are efficient, (ii) verify whether the more
recent \texttt{flexible\_vol} parameters are efficient, and (iii) put the two
policies on the same footing so the operator can compare them on the realised
operational data and choose between them on quantitative grounds rather than on
operational convenience alone.

The objectives induced by Eq.~\eqref{eq:threshold} have no closed form, so we
treat the question as a data-driven decision-making problem
\cite{provost2013} and evaluate it with a discrete-event simulator
\cite{banks2010,law2015} embedded in a simulation-based optimisation loop
\cite{fu2015,amaran2016} over the two-dimensional design space.

\paragraph{Contributions.}
This paper answers the tuning question empirically and makes the following
contributions.
\begin{enumerate}
\item A precise mathematical statement of the dispatch-threshold tuning problem
      as a bi-objective optimisation over $(\beta,\gamma)$, including the
      buffer dynamics, trip-construction rule and the cost and lead-time
      objectives (Section~\ref{sec:model}).
\item A reproducible, data-intensive evaluation pipeline: an event-driven
      simulator calibrated to reproduce the deployed production configuration on
      a full month of real flow, wrapped in a dense grid sweep that
      reconstructs the cost/lead-time Pareto frontier for both policy variants
      (Sections~\ref{sec:method}--\ref{sec:results}).
\item The empirical finding that the deployed \texttt{flexible\_vol}
      configuration is already Pareto-efficient---no evaluated configuration
      dominates it---while the deployed, hand-tuned \texttt{flexible\_shk}
      configuration is Pareto-dominated, with a concrete simulated
      configuration both faster and cheaper, and a cost saving of about
      \num{5.0}\,\pct{} available at equal lead time
      (Section~\ref{sec:results}).
\item A sensitivity analysis establishing that this contrast is robust to grid
      resolution and frontier-reconstruction choices, and a mechanistic
      explanation showing that, given the deployed truck capacity, the
      item-count policy's threshold is mis-tuned: its base is too high and its
      length coupling too small, so a \emph{steeper} threshold is preferable
      (Sections~\ref{sec:sensitivity}--\ref{sec:discussion}).
\item A side-by-side, same-data comparison of the two deployed policy
      variants. Putting both frontiers in the same cost / lead-time plane lets
      the operator pick between \texttt{flexible\_vol} and \texttt{flexible\_shk}
      on a quantitative basis once both signals are available, rather than
      defaulting to the historically convenient choice.
\end{enumerate}

To respect the commercial sensitivity of the operator's data, all monetary
values, cost rates, parcel and trip volumes, and absolute distances and
thresholds are reported here \emph{only} in relative form---as percentages of,
or ratios to, the production reference. The qualitative conclusions do not
depend on the withheld absolute magnitudes.

\section{Related Work}
\label{sec:related}

\paragraph{Vehicle routing and consolidation.}
The classical vehicle-routing problem (VRP) and its many variants ask how to
construct routes that minimise travel cost subject to capacity and time
constraints \cite{toth2014,cordeau2007}. Real last-mile instances rarely admit
the basic VRP form: site-dependent fleets, hard and soft time windows and
split deliveries already push the problem into territory tackled by integer
programming models combined with tailored heuristics
\cite{batsyn2015}, and a network-wide re-routing each period is correspondingly
expensive. Our setting is deliberately narrower and complementary: the routes
are \emph{given} and fixed for the planning period, so the open decision is
temporal consolidation---\emph{when} to release an accumulating load---rather
than spatial routing. This is the dispatch side of the shipment-consolidation
literature surveyed in logistics-systems analysis \cite{daganzo2005}, where
the trade-off between holding goods to fill vehicles and releasing them to
reduce delay is well known. What is specific here is that the operator has
reduced that trade-off to a two-parameter affine rule
(Eq.~\eqref{eq:threshold}) and frozen it into production; we quantify the
price of that simplification on real data.

\paragraph{Multi-objective optimisation.}
Cost and lead time are conflicting objectives, so there is in general no single
optimum but a set of Pareto-efficient policies \cite{miettinen1999}. Population
metaheuristics such as NSGA-II \cite{deb2002} are the standard tool when the
design space is high-dimensional. Our design space, however, is only
two-dimensional, which makes a dense deterministic grid sweep both feasible and
preferable: it reconstructs the frontier directly, without the sampling
variance or convergence assumptions of a metaheuristic, and it makes the
location of a single deployed point relative to the frontier unambiguous.

\paragraph{Simulation and simulation-based optimisation.}
When objectives have no closed form, discrete-event simulation is the standard
evaluation instrument \cite{banks2010,law2015}, and embedding such a simulation
inside an optimisation loop is the subject of a mature literature on
simulation-based optimisation \cite{fu2015,amaran2016}. We adopt exactly this
pattern: the simulator is the objective oracle and the grid sweep is the
optimiser. The contribution is not a new optimisation algorithm but the
application of this data-intensive methodology to audit a deployed policy and to
convert the audit into an actionable reconfiguration.

\paragraph{Shipment consolidation and threshold dispatch.}
The temporal-consolidation decision---whether to dispatch now or wait for more
load---has been studied under quantity-based, time-based and hybrid policies,
where a quantity-based policy releases a shipment once accumulated load reaches a
fixed quantity and a time-based policy releases it after a fixed wait. The
affine rule in Eq.~\eqref{eq:threshold} is a quantity-based policy whose release
quantity is made route-specific through the length term $\gamma\,d_r$. Our
finding---that the length coupling should be close to zero---can be read as
empirical evidence that, for this network, a near-pure quantity policy with a
single shared release level outperforms the length-scaled variant that was
deployed. We are not aware of a published audit that locates a \emph{deployed}
consolidation policy against its empirical Pareto frontier on a complete
operational log, which is the gap this paper fills.

\paragraph{Data-driven decision making.}
Finally, the study sits within the broader programme of data-driven decision
making \cite{provost2013} and data-driven operations
\cite{tao2018}, in which a faithful, data-calibrated model of a running system
is used to test counterfactual policies before deployment. The
production-calibrated simulator plays the role of such a model: it is validated
by reproducing the deployed configuration's behaviour and then used to explore
configurations that were never run in production. The methodological stance is
deliberately conservative---a dense deterministic sweep over an interpretable
two-parameter rule, rather than a black-box optimiser over a richer policy
class---because the deliverable is a configuration change that an operator must
be able to understand, audit and defend.

\section{System Model and Problem Formulation}
\label{sec:model}

\subsection{Network and demand}
Let $\mathcal{R}$ be the set of fixed routes served from a single depot $w$.
Each route $r\in\mathcal{R}$ is an ordered sequence of offices
$(o^{r}_1,\dots,o^{r}_{n_r})$, has a vehicle volume capacity $Q_r$, and a fixed
round-trip distance
\begin{equation}
d_r=\operatorname{dist}(w,o^r_1)+\sum_{i=1}^{n_r-1} \operatorname{dist}(o^r_i,o^r_{i+1})
   +\operatorname{dist}(o^r_{n_r},w),
\label{eq:dr}
\end{equation}
taken from a measured distance matrix. Every office belongs to exactly one
route, so a parcel's destination office determines its route. Demand is a stream
of parcel arrivals
\begin{equation}
\mathcal{B}=\{(a_j,o_j,v_j,s_j)\}_{j=1}^{N},
\end{equation}
where $a_j$ is the arrival time at the depot buffer, $o_j$ the destination
office, $v_j$ the volume, and $s_j$ the item count of parcel $j$. The route of
parcel $j$ is $r(j)$, the unique route whose office set contains $o_j$.

\subsection{Buffer dynamics and dispatch rule}
Each route maintains a buffer of waiting parcels. Define the buffer load under
policy variant $p$ as
\begin{equation}
m^p_r(t)=
\begin{cases}
\dfrac{1}{1000}\displaystyle\sum_{j\in \mathcal{B}_r(t)} v_j & p=\texttt{vol},\\[2.2ex]
\displaystyle\sum_{j\in \mathcal{B}_r(t)} s_j & p=\texttt{shk},
\end{cases}
\label{eq:load}
\end{equation}
where $\mathcal{B}_r(t)$ is the set of parcels for route $r$ present in the
buffer at time $t$; in the \texttt{vol} case volumes are converted to litres.
With design parameters $\theta=(\beta,\gamma)$, a dispatch is triggered for route
$r$ at the first arrival epoch at which
\begin{equation}
m^p_r(t)\;\ge\;\tau_r(\theta)=\beta+\gamma\,d_r .
\label{eq:trigger}
\end{equation}
The same scalar pair $(\beta,\gamma)$ governs every route; the per-route
heterogeneity enters only through $d_r$. The rule is intentionally minimal,
which is precisely why a single re-tuning can have network-wide effect.

\subsection{Trip construction}
On a dispatch at time $t$, parcels are loaded greedily, in arrival order, while
the cumulative loaded volume stays below a fill ceiling $\rho\,Q_r$ (with
$\rho=0.8$); any parcels that do not fit remain in the buffer for the next trip.
The vehicle visits only the offices that received at least one loaded parcel, in
their fixed route order. After a constant loading delay $h_L$, travel times
between consecutive stops are taken from a measured duration matrix, and the
dwell at an office holding $k$ parcels is $u_0+u_1 k$. Let $D_k$ and $T_k$ denote
the distance and total elapsed time of trip $k$, and let
\begin{equation}
\ell_j=\big(\text{arrival time of the serving vehicle at }o_j\big)-a_j
\label{eq:ellj}
\end{equation}
be the \emph{lead time} of parcel $j$, measured from buffer entry to office
arrival.

\subsection{Objectives}
Over the simulation horizon the company operating cost is
\begin{equation}
C(\theta)=c_{\mathrm{km}}\sum_{k} D_k
         +c_{\mathrm{h}}\sum_{k}\frac{T_k}{3600},
\qquad
\kappa=\mu\,\eta\,\sigma,
\label{eq:cost}
\end{equation}
where $c_{\mathrm{km}}$ is the cost per kilometre, and $c_{\mathrm{h}}$ the cost per
vehicle hour. In the calibrated data this credit is below
\num{0.1}\,\pct{} of $C$, so cost is governed almost entirely by accumulated
distance and driver hours; both grow with the number of trips, which is exactly
what the dispatch threshold controls. The average parcel lead time is
\begin{equation}
L(\theta)=\frac{1}{N}\sum_{j=1}^{N}\ell_j .
\label{eq:lead}
\end{equation}

\subsection{Bi-objective problem}
Both objectives are to be minimised. The design problem is
\begin{equation}
\min_{\theta=(\beta,\gamma)\in\Theta}\;\;\big(C(\theta),\,L(\theta)\big).
\label{eq:biobj}
\end{equation}
A configuration $\theta'$ \emph{dominates} $\theta$ if $C(\theta')\le C(\theta)$
and $L(\theta')\le L(\theta)$ with at least one strict inequality. The
\emph{Pareto frontier}
\begin{equation}
\mathcal{P}=\{\theta\in\Theta:\nexists\,\theta'\in\Theta\ \text{dominating}\ \theta\}
\end{equation}
collects the efficient policies. The production configuration $\theta_0$ is a
single point in $\Theta$; the empirical question of this paper is whether
$\theta_0\in\mathcal{P}$, and if not, how far inside the frontier it lies and in
which direction the efficient configurations differ from it.

\section{Method and Experimental Setup}
\label{sec:method}

\subsection{Discrete-event simulator}
$C(\theta)$ and $L(\theta)$ have no closed form: they emerge from the
interaction of stochastic arrivals, capacity-limited packing, and fixed
geometry. We therefore evaluate them with an event-driven simulator that replays
the real arrival stream in timestamp order, maintains the per-route buffers,
applies the trigger \eqref{eq:trigger}, constructs trips exactly as described in
Section~\ref{sec:model}, and aggregates \eqref{eq:cost}--\eqref{eq:lead} over the
horizon. Each evaluation of $F(\theta)=(C(\theta),L(\theta))$ is one full replay
of the month. The simulator is deterministic given the recorded arrival stream
and a fixed $\theta$, so repeated evaluations are exactly reproducible---a
prerequisite for an unambiguous Pareto comparison.

\subsection{Data pipeline}
The simulator is driven by one month of production data: the full real arrival
stream across the network's routes and offices, with measured inter-office
distances and durations supplied as matrices. The ingestion pipeline joins three
sources---the arrival log (timestamp, destination office, volume, item count per
parcel), the route definitions (ordered office sequences and vehicle
capacities), and the measured distance/duration matrices---into the event stream
consumed by the simulator. The cost coefficients $c_{\mathrm{km}},c_{\mathrm{h}}$
and the production design points
$\theta_0^{\texttt{vol}},\theta_0^{\texttt{shk}}$ are the deployed values. No
synthetic demand is used; the study is a replay of recorded operations.

\subsection{Validation against the production reference}
Before sweeping the design space we validate the model by running it at the
deployed configuration $\theta_0$ and confirming that it reproduces the
production reference operating point---the realised cost and average lead time of
the live policy. This calibration step anchors the relative comparisons that
follow: every percentage in Section~\ref{sec:results} is expressed against this
reproduced reference, which is fixed at $(100\,\pct,100\,\pct)$ by construction.

\subsection{Frontier reconstruction}
Because $\theta$ is two-dimensional, the frontier is recovered by a dense grid
sweep rather than a metaheuristic. We evaluate $F$ on a Cartesian grid
$\{\beta_i\}\times\{\gamma_j\}$, mark the non-dominated points by pairwise
comparison, and connect them, ordered by cost, into a piecewise-linear frontier.
Linear interpolation along the frontier then yields, for any target lead time,
the minimum achievable cost and the $(\beta,\gamma)$ that attains it---exactly
the comparison we make against the production point. Grid resolutions are listed
in Table~\ref{tab:setup}; the evaluations are independent and were executed in
parallel. For \texttt{flexible\_vol} the sweep used \num{136} evaluations
(\num{17} levels of $\beta$ by \num{8} levels of $\gamma$) and for
\texttt{flexible\_shk} \num{117} evaluations (\num{13}$\times$\num{9}); together
with the two production reference points this gives \num{255} full-month replays
in total. Procedure~\ref{proc:front} states the reconstruction precisely.

\begin{figure}[t]
\centering
\fbox{\begin{minipage}{0.93\linewidth}
\small
\textbf{Procedure 1: Pareto-frontier reconstruction.}\\[2pt]
\emph{Input:} grid $G=\{(\beta_i,\gamma_j)\}$; simulator $F(\theta)=(C,L)$;
production point $\theta_0$ with lead time $L_0$.\\[2pt]
\emph{Output:} efficient frontier $\mathcal{P}$; minimum cost
$C^\star$ at lead time $L_0$.
\begin{enumerate}\setlength{\itemsep}{1pt}
\item For every $\theta\in G$, evaluate $(C(\theta),L(\theta))$ by one full-month
      replay (independent, run in parallel).
\item Mark $\theta$ \emph{non-dominated} if no other evaluated $\theta'$ satisfies
      $C(\theta')\le C(\theta)$ and $L(\theta')\le L(\theta)$ with one strict
      inequality; collect the non-dominated points into $\mathcal{P}$.
\item Sort $\mathcal{P}$ by cost and connect consecutive points into a
      piecewise-linear curve.
\item Read $C^\star$ off this curve at lead time $L_0$ by linear interpolation;
      the saving at equal lead time is $1-C^\star/C(\theta_0)$.
\end{enumerate}
\end{minipage}}
\caption{Deterministic frontier reconstruction. Steps~2--4 only ever remove
points, so the reported saving is a lower bound at the grid's resolution.}
\label{proc:front}
\end{figure}

\begin{table}[t]
\centering
\caption{Simulation inputs and grid-sweep configuration (relative form).}
\label{tab:setup}
\begin{tabular}{lll}
\toprule
\multicolumn{3}{l}{\emph{Network and demand}}\\
\midrule
Depot               & \multicolumn{2}{l}{single}\\
Horizon             & \multicolumn{2}{l}{$\sim$1 month of production flow}\\
Fill ceiling $\rho$ & \multicolumn{2}{l}{0.8}\\
Demand              & \multicolumn{2}{l}{real recorded arrival stream (no synthetic data)}\\
\midrule
\multicolumn{3}{l}{\emph{Design grid} $(\beta,\gamma)$, number of levels}\\
\midrule
                   & \texttt{flexible\_vol} & \texttt{flexible\_shk}\\
$\beta$ levels     & 17 & 13\\
$\gamma$ levels    & 8  & 9\\
evaluations        & 136 & 117\\
\bottomrule
\end{tabular}
\end{table}

\subsection{Computational cost and reproducibility}
The audit is cheap. A single evaluation is one deterministic replay of the
recorded month; the full study is the \num{255} replays of the two grids, which
are mutually independent and embarrassingly parallel. Because the simulator is
deterministic given the recorded arrival stream and a fixed $\theta$, the entire
experiment is bit-for-bit reproducible: re-running the sweep regenerates the same
frontier, and the production reference is recovered exactly each time the
deployed $\theta_0$ is replayed. We persist, for each evaluated $\theta$, the
pair $(C,L)$ together with the resolved $(\beta,\gamma)$, so the frontier and all
relative figures in this paper can be regenerated from the stored grid without
re-running the simulator---an important property for an audit that an operator
may wish to repeat on a later month.

\section{Results}
\label{sec:results}

\subsection{The volume policy is already Pareto-efficient}
\begin{figure}[t]
\centering
\begin{subfigure}{0.49\linewidth}
\centering
\includegraphics[width=\linewidth]{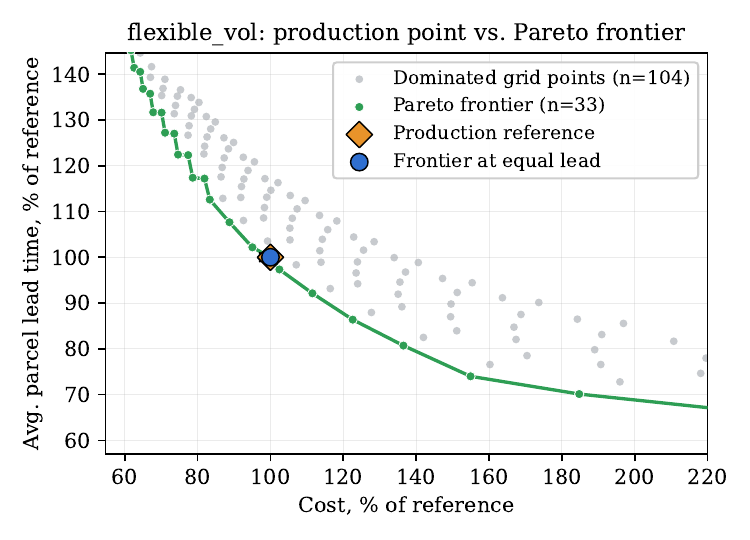}
\caption{\texttt{flexible\_vol}: production point sits on the frontier;
no improvement available.}
\label{fig:vol}
\end{subfigure}
\hfill
\begin{subfigure}{0.49\linewidth}
\centering
\includegraphics[width=\linewidth]{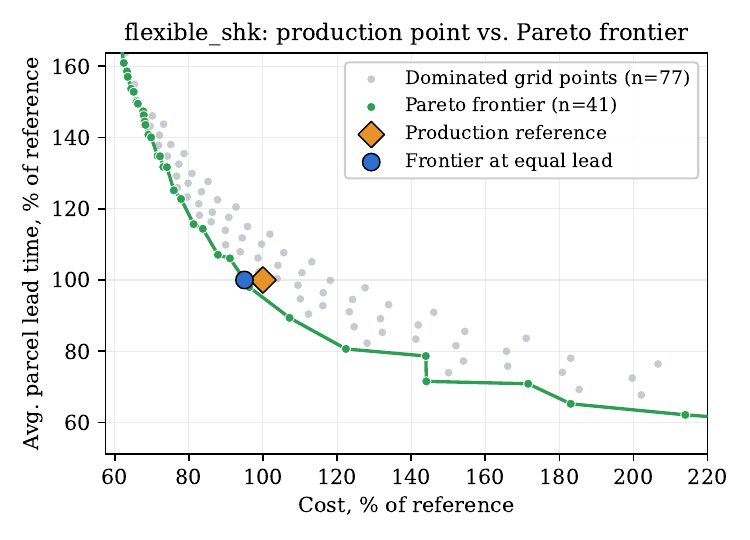}
\caption{\texttt{flexible\_shk}: production point dominated; \num{5.0}\,\pct{}
available at equal lead.}
\label{fig:shk}
\end{subfigure}
\caption{Reconstructed cost / lead-time Pareto frontiers, with cost and lead
time shown as a percentage of the production reference ($=100\,\pct$). Grey:
dominated grid configurations; green: efficient frontier; orange diamond:
production configuration; blue: frontier point at equal lead time.}
\label{fig:fronts}
\end{figure}

Figure~\ref{fig:vol} shows the reconstructed frontier for the
\texttt{flexible\_vol} policy, with both axes expressed as a percentage of the
production reference. The grey points are dominated grid configurations; the
green curve is the efficient frontier; the orange diamond is the production
configuration $\theta_0^{\texttt{vol}}$, fixed at $(100\,\pct,100\,\pct)$ by
construction. The production point sits \emph{on} the frontier: it is
non-dominated within the evaluated grid, and the script's interpolation at the
production lead time returns the production cost itself
(Table~\ref{tab:results}). No evaluated $(\beta,\gamma)$ pair is both cheaper
and no slower; in particular, neither lead reduction at fixed cost nor cost
reduction at fixed lead time is available at the grid's resolution.

This is the answer an operator hopes the audit will return. The production
point sits at the frontier's knee: pushing $(\beta,\gamma)$ toward shorter
lead times costs disproportionately more, while relaxing it would trade speed
for cost---a stable operating regime with no re-tuning available.

\subsection{The item-count policy is Pareto-dominated}
Figure~\ref{fig:shk} repeats the analysis for \texttt{flexible\_shk}. Here the
production point lies visibly above and to the right of the frontier and is
\emph{dominated} by a concrete simulated configuration that is both cheaper and
faster. Quantitatively (Table~\ref{tab:results}), at the production lead time
the frontier attains the same service at \num{5.0}\,\pct{} lower cost.
Equivalently, holding cost fixed at the production level, the frontier reaches
a lead time \num{4.9}\,\pct{} shorter. The dominating evaluated configuration
uses a lower base and a substantially steeper length coupling than the deployed
policy---we return to this mechanism in Section~\ref{sec:discussion}.

The two policies are therefore in markedly different states of tune. They also
occupy similar but not identical operating regimes: in absolute terms the
item-count policy runs at about \num{83}\,\pct{} of the volume policy's cost
and \num{1.19}$\times$ its lead time. Neither policy dominates the other across
the whole range---an operator with a strong cost preference would deploy
\texttt{flexible\_shk}, an operator with a stronger speed preference would
deploy \texttt{flexible\_vol}---so the \num{5.0}\,\pct{} inefficiency in
\texttt{flexible\_shk} matters: it is paid every time the cheaper policy is
chosen.

\begin{table}[t]
\centering
\caption{Production reference vs.\ efficient frontier, all quantities relative
to the production reference (comparison at equal lead time).}
\label{tab:results}
\begin{tabular}{lcccc}
\toprule
& \multicolumn{2}{c}{\texttt{flexible\_vol}} & \multicolumn{2}{c}{\texttt{flexible\_shk}}\\
\cmidrule(lr){2-3}\cmidrule(lr){4-5}
& Reference & Frontier & Reference & Frontier\\
\midrule
Base $\beta$ (rel.)             & $1.00\times$ & $1.00\times$ & $1.00\times$ & $0.84\times$\\
Length coupling $\gamma$ (rel.) & $100\,\pct$ & $100\,\pct$
                                & $100\,\pct$ & $157\,\pct$\\
Cost (rel.)                     & $100\,\pct$ & $100\,\pct$
                                & $100\,\pct$ & $95.0\,\pct$\\
Lead time (rel.)                & $100\,\pct$ & $100\,\pct$
                                & $100\,\pct$ & $100\,\pct$\\
\midrule
Cost saving at equal lead       & \multicolumn{2}{c}{$0\,\pct$ (already on frontier)}
                                & \multicolumn{2}{c}{$5.0\,\pct$}\\
\bottomrule
\end{tabular}
\end{table}

\section{Sensitivity and Robustness}
\label{sec:sensitivity}

A single grid sweep raises three natural objections: that the
\texttt{flexible\_shk} dominance gap is an artefact of grid resolution, that it
depends on the interpolation used to read the frontier at the production lead
time, or that it is peculiar to the exact lead-time target chosen. We address
each in turn, and note in passing how the same arguments support the parallel
conclusion that \texttt{flexible\_vol} is on its frontier.

\paragraph{Grid resolution.}
The frontier is reconstructed from the non-dominated subset of the evaluated
grid (\num{33} of \num{137} points are non-dominated for \texttt{flexible\_vol},
\num{41} of \num{118} for \texttt{flexible\_shk}). Because dominance is a
pairwise relation that only ever \emph{removes} points, refining the grid can
only push the reconstructed frontier further from a dominated production point
or leave it unchanged---it can never make a dominated point efficient. The
reported \num{5.0}\,\pct{} \texttt{flexible\_shk} saving is therefore a
conservative lower bound at the grid's resolution; a finer grid would weakly
increase it. The complementary statement for \texttt{flexible\_vol} is weaker:
the volume policy's production point is non-dominated within the evaluated grid,
so a finer grid could in principle reveal a near-by dominating configuration we
have not yet sampled. With the present grid spacing, however, no such point
exists, and the basin of evaluated points around the volume reference is
uniformly worse on at least one objective.

\paragraph{Interpolation.}
The equal-lead comparison reads the frontier cost at the production lead time by
linear interpolation between adjacent non-dominated points. Since the frontier
is convex-like and piecewise linear, linear interpolation between two efficient
points yields a cost no lower than the true frontier, so the interpolated saving
again \emph{understates} the gap. Moreover, the \texttt{flexible\_shk}
dominance conclusion does not rely on interpolation at all: a concrete
evaluated grid configuration already dominates the item-count production point
on both objectives simultaneously, so the result holds even if the interpolated
frontier is discarded entirely.

\paragraph{Choice of operating point.}
The \num{5.0}\,\pct{} \texttt{flexible\_shk} figure is quoted at the production
lead time, but the production point is dominated over a neighbourhood, not only
at that single target: the gap can be expressed equivalently as a
\num{4.9}\,\pct{} lead-time reduction at fixed cost, and the dominating
evaluated configuration improves both objectives at once. The conclusion---that
$\theta_0^{\texttt{shk}}\notin\mathcal{P}^{\texttt{shk}}$ while
$\theta_0^{\texttt{vol}}\in\mathcal{P}^{\texttt{vol}}$ at the current
resolution---is therefore insensitive to which of the two objectives is held
fixed.

\paragraph{What would shift the result.}
The simulator assumes unlimited vehicles and deterministic travel times.
Finite fleets or congestion would move both frontiers as a whole but, as the
inefficiency is structural (Section~\ref{sec:discussion}), would not
selectively rescue the dominated item-count point. The relative positions
---volume on its frontier, item-count off it---should persist.

\section{Discussion: why the item-count policy is mis-tuned}
\label{sec:discussion}

\begin{figure}[tb]
\centering
\includegraphics[width=0.74\linewidth]{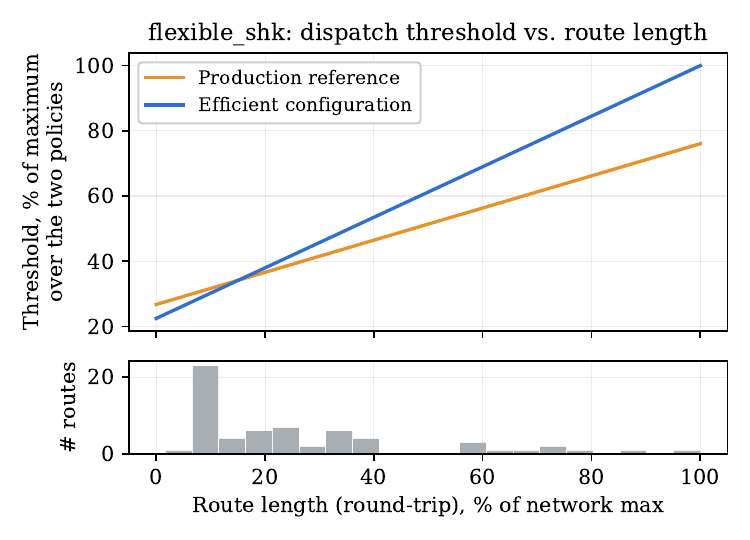}
\caption{Dispatch threshold $\tau_r=\beta+\gamma d_r$ for the production and
efficient \texttt{flexible\_shk} policies, normalised: the threshold is shown as
a percentage of the larger of the two threshold values at the longest route,
and route length as a percentage of the network maximum (lower panel:
route-length distribution). The production rule's shallow slope makes the
threshold weakly length-dependent; the efficient rule rises more sharply with
route length.}
\label{fig:thresh}
\end{figure}

The frontier parameters explain the \texttt{flexible\_shk} gap. The production
item-count policy uses a relatively large base $\beta$ and a relatively small
length coefficient $\gamma$, so its threshold is nearly flat across route
lengths. The efficient configuration moves in the opposite direction: a lower
base and a substantially larger length coupling ($\beta$ drops to about
\num{0.84}$\times$ the production value, while $\gamma$ rises to about
\num{157}\,\pct{}---Table~\ref{tab:results}), making the threshold visibly
\emph{steeper} as a function of route length.
Figure~\ref{fig:thresh} draws both threshold lines, normalised, over the
distribution of route lengths.

The shallow production rule is sub-optimal at both ends of the length
distribution. For the many short routes (lower panel of
Figure~\ref{fig:thresh}), the production threshold is set higher than the
efficient steep rule prescribes, so loads accumulate for too long before a
truck is released; that inflates customer lead time on short routes without
buying a meaningful cost reduction. For the few very long routes the situation
reverses: with the deployed \num{20}\,m$^3$ trucks there is plenty of unused
capacity at the production threshold, and the efficient rule pushes the
threshold higher so more parcels travel per trip. The cost contribution from
those long routes (per kilometre and per vehicle hour) is large, so packing
more into each long-route trip is exactly where the \num{5.0}\,\pct{} cost
saving comes from.

The volume policy meters dispatch directly in the same units as truck
capacity, so its affine threshold tracks capacity for free; the item-count
policy must track capacity indirectly through $(\beta,\gamma)$, and the
deployed pair does this poorly. A steeper threshold restores the relationship.

\paragraph{Practical implication.}
The remedy requires no change to the routing, the fleet, or the software---only
two numbers in the item-count configuration. Replacing the production pair with
a lower-base, higher-coupling threshold is a zero-capital intervention that the
simulation projects to cut \texttt{flexible\_shk} cost by about
\num{5.0}\,\pct{}, an actionable recurring saving. Because the simulator
reproduces the production reference, the operator can audit the proposed
configuration in advance by replaying the same month under the new parameters
before committing them to production. The volume policy needs no such change;
the audit's positive verdict on \texttt{flexible\_vol} is itself useful, as it
removes one production parameter from the list of candidates for re-tuning.

\paragraph{Limitations.}
The study covers one month of operational flow at one regional warehouse, with
one truck capacity, so the quantitative figures are demand-, region- and
fleet-specific; in particular, the same \texttt{flexible\_shk} audit run
against a different truck size would in general find a different efficient
$(\beta,\gamma)$, and could even shift the volume-policy verdict. The
operator's actual network comprises more than one hundred logistic centres
spread across many regions, each with its own route topology, route-length
distribution and demand mix, so the efficient $(\beta,\gamma)$ found here is
not expected to be efficient for every region---an opportunity we return to in
Section~\ref{sec:conclusion}. The qualitative result---that the item-count
policy's efficient threshold is steeper than the deployed one given the
present truck capacity---should be re-validated whenever capacity or cost
coefficients change. The simulator also assumes unlimited vehicles and
deterministic travel times, so queueing or congestion would shift the
frontier, though, as argued in Section~\ref{sec:sensitivity}, not the relative
positions of the two production points. Finally, the dispatch rule is fixed to
the affine form in Eq.~\eqref{eq:threshold}; richer rules (e.g.\ time-of-day
or per-route thresholds) might extend the frontier further, but evaluating
them is outside the scope of auditing the deployed policy.

\section{Conclusion}
\label{sec:conclusion}

We formalised the fixed-route dispatch decision as a two-parameter,
bi-objective tuning problem and reconstructed its cost/lead-time Pareto frontier
by direct simulation on a month of real flow, treating the deployed policy as
the object of a data-driven audit. The two policy variants in production receive
opposite verdicts. The volume-threshold configuration sits on its frontier: no
evaluated $(\beta,\gamma)$ pair dominates it, and the audit therefore returns
the unusual positive result that this policy is already efficient under the
current cost coefficients and deployed truck capacity. The item-count
configuration is the opposite: it is Pareto-dominated by a concrete simulated
configuration that is both faster and cheaper, with about \num{5.0}\,\pct{} of
cost recoverable at equal lead time. The mechanism, given the deployed truck
size, is a base that is too high and a length coefficient that is too small;
the efficient configuration uses a lower base and a steeper slope so that long
routes pack more parcels per trip and short routes dispatch sooner.
Because the fix is a two-scalar reconfiguration that can be validated on the
same production data before deployment, the analysis converts directly into an
actionable, zero-capital saving on the item-count policy and a verified clean
bill on the volume policy. Beyond these two verdicts, the simulator supports a
third decision: which of the two threshold rules to deploy at all. Historically
\texttt{flexible\_shk} was used because the item-count signal is recorded
exactly while the volume signal carries warehouse-management measurement
noise; once both signals are available, the simulator removes the
convenience-vs.-optimality conflict by placing both frontiers in the same
plane. The simulator therefore plays three complementary roles---re-tuning the
mis-tuned hand-crafted item-count policy, confirming the more recent volume
policy efficient, and supporting a principled choice between them---at the
cost of a few hundred deterministic replays of the recorded month.

\paragraph{Scaling the audit network-wide.}
The audit reported here was run on the operational flow of a single regional
warehouse, but the operator runs a network of more than one hundred logistic
centres across many regions. Each region has its own route topology, its own
distribution of route lengths and its own demand mix, so the efficient
$(\beta,\gamma)$ found in one region is not in general efficient in another:
no single, network-wide setting of the two scalars can be on every region's
frontier at once. The audit procedure, by contrast, is fully driven by the
per-warehouse operational log and takes the same inputs as the production
configuration itself, so it can be re-run independently for every region to
recover a region-specific efficient $(\beta,\gamma)$. A network-wide rollout
therefore turns the single-region cost reduction reported here into a
recurring saving that scales with the number of regions audited---an effect
that a uniform central re-tuning cannot capture, since the efficient choice is
demand- and route-structure-specific. Region-by-region tuning of the
dispatch rule is, in this sense, the natural production-scale extension of
the present audit, and the principal business case for embedding the simulator
in the operator's planning loop.

\subsubsection*{Acknowledgments.}
The study was implemented in the framework of the Basic Research Program at
HSE University (HSE-BR-2025-080).

\bibliographystyle{splncs04}
\bibliography{public_paper_lncs}

\end{document}